# Prospects for the CERN Axion Solar Telescope Sensitivity to 14.4 keV Axions

K. Jakovčić,[a, *] S. Andriamonje,[c] S. Aune,[c] F. Avignone,[d] K. Barth,[b] A. Belov,[l]
B. Beltrán,[g,1] H. Bräuninger,[f] J. M. Carmona,[g] S. Cebrián,[g] J. I. Collar,[h] T. Dafni,[e,2]
M. Davenport,[b] L. Di Lella,[b,3] C. Eleftheriadis,[i] G. Fanourakis,[j] E. Ferrer-Ribas,[c]
H. Fischer,[k] J. Franz,[k] P. Friedrich,[f] T. Geralis,[j] I. Giomataris,[c] S. Gninenko,[l]
M. D. Hasinoff,[m] F. H. Heinsius,[k] D. H. H. Hoffmann,[e] I. G. Irastorza,[c] J. Jacoby,[n]
D. Kang,[k] K. Königsmann,[k] R. Kotthaus,[o] M. Krčmar,[a] K. Kousouris,[j] M. Kuster,[e,f]
B. Lakić,[a] C. Lasseur,[b] A. Liolios,[i] A. Ljubičić,[a] G. Lutz,[o] G. Luzón,[g] D. W. Miller,[h]
A. Morales,[g,4] J. Morales,[g] A. Ortiz,[g] T. Papaevangelou,[b] A. Placci,[b] G. Raffelt,[o]
J. Ruz,[g] H. Riege,[e] Y. Semertzidis,[p,5] P. Serpico,[o] L. Stewart,[b] J. D. Vieira,[h] J. Villar,[g]
J. Vogel,[k] L. Walckiers,[b] K. Zioutas[p]
(CAST Collaboration)

[a]*Ruđer Bošković Institute, Bijenička cesta 54, P.O.Box 180, HR-10002 Zagreb, Croatia*

[b]*European Organization for Nuclear Research (CERN), CH-1211 Genève 23, Switzerland*

[c]*DAPNIA, Centre d'Études Nucléaires de Saclay (CEA-Saclay), Gif-sur-Yvette, France*

[d]*Department of Physics and Astronomy, University of South Carolina, Columbia, SC, USA*

[e]*Gesellschaft für Schwerionenforschung, GSI-Darmstadt, Plasmaphysik, Planckstr. 1, 64291 Darmstadt, and Technische Universität Darmstadt, IKP, Schlossgartenstrasse 9, 64289 Darmstadt, Germany*

[f]*Max-Planck-Institut für extraterrestrische Physik, Garching, Germany*

[g]*Instituto de Física Nuclear y Altas Energías, Universidad de Zaragoza, Zaragoza, Spain*

[h]*Enrico Fermi Institute and KICP, University of Chicago, Chicago, IL, USA*

---

[*] Corresponding author. Tel.:+385-1-4560957; fax:+385-1-4680239; e-mail: Kresimir.Jakovcic@irb.hr
[1] Present address: Department of Physics, Queen's University, Kingston, Ontario K7L 3N6
[2] Present address: DAPNIA, CEA-Saclay, Gif-sur-Yvette, France
[3] Present address: Scuola Normale Superiore, Pisa, Italy
[4] Deceased
[5] Permanent address: Brookhaven National Laboratory, NY-USA

[i]*Aristotle University of Thessaloniki, Thessaloniki, Greece*
[j]*National Center for Scientific Research "Demokritos", Athens, Greece*
[k]*Albert-Ludwigs-Universität Freiburg, Freiburg, Germany*
[l]*Institute for Nuclear Research (INR), Russian Academy of Sciences, Moscow, Russia*
[m]*Department of Physics and Astronomy, University of British Columbia, Vancouver, Canada*
[n]*Johann Wolfgang Goethe-Univ., Inst. für Angewandte Physik, Frankfurt am Main, Germany*
[o]*Max-Planck-Institut für Physik (Werner-Heisenberg-Institut), Munich, Germany*
[p]*Physics Department, University of Patras, Patras, Greece*

**Abstract**

The CERN Axion Solar Telescope (CAST) is searching for solar axions using the 9.0 T strong and 9.26 m long transverse magnetic field of a twin aperture LHC test magnet, where axions could be converted into X-rays via reverse Primakoff process. Here we explore the potential of CAST to search for 14.4 keV axions that could be emitted from the Sun in M1 nuclear transition between the first, thermally excited state, and the ground state of $^{57}$Fe nuclide. Calculations of the expected signals, with respect to the axion-photon coupling, axion-nucleon coupling and axion mass, are presented in comparison with the experimental sensitivity.

## 1. Introduction

Axions are hypothetical neutral pseudoscalar particles that arise in the context of the Peccei-Quinn solution [1,2] to the strong CP problem associated with the Θ-vacuum structure of QCD. Due to their potential abundance in the early Universe, axions are also viable dark matter candidates. As axions couple to photons and nucleons, the Sun would be a strong axion emitter. Axions of a broad energy spectrum with an average energy of 4.2 keV could be produced abundantly in the core of the Sun by the Primakoff conversion of thermal photons in the fluctuating electromagnetic fields of nuclei and electrons in the solar plasma. Furthermore, axion emission from the magnetic nuclear transitions in some nuclides that are present in the Sun would provide an additional component to solar axions. Axions produced by the latter mechanism would be monoenergetic since their energy corresponds to the energy of the particular nuclear transition. Some nuclear processes have been proposed as sources of solar monoenergetic axions [3,4] and experiments based on the detection of the axions with suppressed electromagnetic couplings were reported by several authors [5-8]. Here we focus on the case when the (solar) hadronic axions couple to nucleons and to two photons.

## 2. CAST experiment

One method of detecting solar axions on the Earth is to make the Primakoff effect work in reverse, using a strong transverse magnetic field in a laboratory to coherently convert the solar axions into X-ray photons. Inside the magnetic field, the axion couples to a virtual photon producing a real photon: $a+\gamma_{virtual} \rightarrow \gamma$. The energy of this photon is equal to the total energy of the axion. The CERN Axion Solar Telescope (CAST) [9] is primarily designed to search for solar axions generated via the Primakoff process using the above mentioned method. The magnetic field in the CAST experiment is provided by a decommissioned Large Hadron Collider (LHC) prototype dipole magnet, which produces a magnetic field of 9.0 T in the interior of two parallel, straight, 9.26 m long pipes with a cross sectional area $A = 2 \times 14.5$ cm$^2$. The magnet is mounted on a moving platform with $\pm 8°$ vertical and $\pm 40°$ horizontal movement, allowing observation of the Sun for 1.5 h at both sunrise and sunset during the whole year. The time when the Sun is not reachable is devoted to background measurements. At the both ends of the magnet, low background X-ray detectors are installed to search for photons coming from axion conversion in the magnet when it is pointing to the Sun. The analysis of the first CAST data has established an upper limit to the axion-photon coupling of $g_{a\gamma\gamma} < 1.16 \times 10^{-10}$ GeV$^{-1}$ at 95% C.L. for axion mass $m_a \lesssim 0.02$ eV [10]. This limit improves the best previous laboratory constraints on $g_{a\gamma\gamma}$ by a factor 5 and is comparable to the limit derived from stellar energy-loss arguments.

## 3. Estimation of the CAST sensitivity to 14.4 keV solar axions

Here we explore the potential of the CAST to search for 14.4 keV axions that could be emitted from the Sun, in M1 nuclear transition between the first, thermally excited state, and the ground state of $^{57}$Fe nuclide [4]. $^{57}$Fe can be a suitable axion emitter due to its exceptional abundance among the heavy elements in the Sun (solar abundance by mass fraction $2.7 \times 10^{-5}$), and its first excitation energy of 14.4 keV is low enough to be excited thermally in the hot solar interior. Following the calculations in Ref. [11] and Ref. [12], the expected axion flux at the Earth due to the $^{57}$Fe de-excitations in the Sun is

$$\Phi_a = 4.405 \times 10^{23} (-1.19 g_0 + g_3)^2 \text{ cm}^{-2}\text{s}^{-1}, \quad (1)$$

where $g_0$ and $g_3$ are the isoscalar and isovector axion-nucleon coupling constants and they are model dependent parameters. For example, $g_0$ and $g_3$ for KSVZ (Kim-Shifman-Vainshtein-Zakharov) or hadronic axions [13,14] are defined as

$$g_0 = -\frac{m_N}{f_{PQ}} \frac{1}{6} \left[ 2S + (3F - D) \frac{1 + z - 2w}{1 + z + w} \right] \quad (2)$$

and

$$g_3 = -\frac{m_N}{f_{PQ}} \frac{1}{2} (D + F) \frac{1 - z}{1 + z + w}, \quad (3)$$

where $m_N$ is the nucleon mass, $f_{PQ}$ denotes the Peccei-Quinn symmetry scale, while $z \equiv m_u / m_d \approx 0.56$ and $w \equiv m_u / m_s \approx 0.029$ are the quark mass ratios. The constants $F = 0.46$ and $D = 0.8$ are the invariant matrix elements for the SU(3) octet axial vector currents and $S \approx 0.58$ characterizes the flavor singlet coupling.

The expected number of X-rays produced by the conversion of 14.4 keV solar axions in the CAST magnet is

$$N_\gamma = \Phi_a P_{a \to \gamma} A t, \quad (4)$$

where $t$ is the measurement time and $P_{a \to \gamma}$ is the axion-photon conversion probability. This probability in vacuum is

$$P_{a \to \gamma} = \left( \frac{g_{a\gamma\gamma} B}{q} \right)^2 \sin^2 \left( \frac{qL}{2} \right), \quad (5)$$

where $B$ is the magnetic field strength, $L$ is its length, while $q = m_a^2 / (2 E_a)$ is the axion-photon momentum difference. Using these relations, in case that axion signal is not detected, one can estimate the upper limit to the axion-photon coupling as a function of axion mass and axion-nucleon coupling:

$$\frac{g_{a\gamma\gamma}}{10^{-10} \text{GeV}^{-1}} \leq \frac{q}{(-1.19 g_0 + g_3) B} \times \frac{1}{\sin(qL/2)} \sqrt{\frac{3 \sqrt{N_B}}{4.405 \times 10^3 A t \varepsilon}}. \quad (6)$$

Here $N_B$ is the number of background events in the detector and $\varepsilon$ is the detection efficiency.

## 4. Results and discussion

For our calculations, we assumed two-year experiment run (~ 33 days of integrated exposure to the Sun). The corresponding estimated background is $N_B = 2145$ counts, while the detection efficiency for the converted X-rays is assumed to be 100% [9]. In Fig. 1 we present our estimation of the $3\sigma$ level upper limit to $g_{a\gamma\gamma}$ as a function of $m_a$ for three different axion-nucleon couplings. In the middle contour we have taken $(-1.19g_0 + g_3) = 1.55 \times 10^{-7}$, the KSVZ axion model value for the Peccei-Quinn symmetry breaking scale $f_{PQ} = 10^6$ GeV. The other two contours correspond to fixing $(-1.19g_0 + g_3)$ at 1/10 and 10 times this value. One can thus easily scale the excluded region for any choice of $g_0$ and $g_3$. Therefore, our result can also be generally applied (i.e. valid for general models incorporating axions) to impose the constraints for any light pseudoscalars that couple to the nucleon and to two photons.

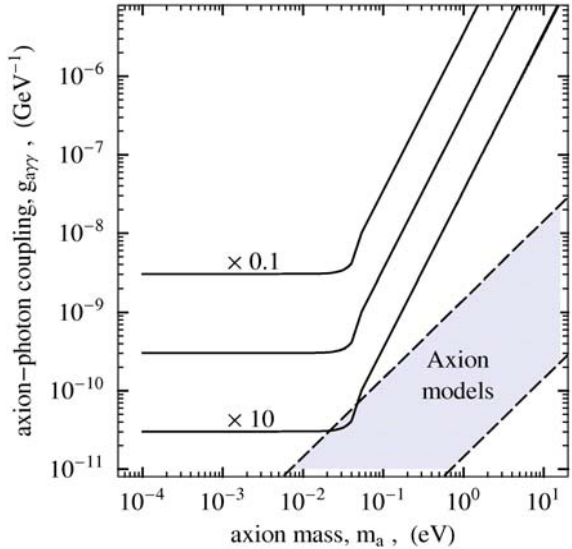

Fig. 1. Upper limits imposed on $g_{a\gamma\gamma}$ by the estimation of the CAST sensitivity to 14.4 keV solar axions are given as a function of $m_a$. The $3\sigma$ limits are shown for three axion nuclear coupling values indicated in the text. The shaded band represents typical theoretical models.

## Acknowledgments

We thank CERN for hosting the experiment. We acknowledge support from NSERC (Canada), MSES (Croatia), CEA (France), BMBF (Germany) GSRT (Greece), RFFR (Russia), CICyT (Spain), and NSF (USA). We acknowledge the helpful discussions within the network on direct dark matter detection of the ILIAS integrating activity (Contract number: RII3-CT-2003-506222). This project was also supported by the Bundesministerium für Bildung und Forschung (BMBF) under the grant number 05 CC2EEA/9 and 05 CC1RD1/0, and by the Virtuelles Institut für Dunkle Materie und Neutrinos – VIDMAN.